\documentclass[letterpaper,aps,prd,preprint,nofootinbib,superscriptaddress]{revtex4}
\pdfoutput=0
\usepackage{epsf}
\usepackage{epsfig}
\usepackage{graphics}
\usepackage{graphicx}
\usepackage{amssymb}
\usepackage{amsmath}
\newcommand{\nc}{\newcommand}
\nc{\beq}{\begin{equation}}   \nc{\eeq}{\end{equation}}
\nc{\bea}{\begin{eqnarray}}   \nc{\eea}{\end{eqnarray}}
\nc{\baa}{\begin{array}}      \nc{\eaa}{\end{array}}
\nc{\bit}{\begin{itemize}}    \nc{\eit}{\end{itemize}}
\nc{\ben}{\begin{enumerate}}  \nc{\een}{\end{enumerate}}
\nc{\bce}{\begin{center}}     \nc{\ece}{\end{center}}

\begin{document}

%% \title{Flavor Bounds and Phenomenology in the Scalar Sector of RS
%%   Scenarios\footnote{Talk presented at the SUSY09 Conference, Northeastern
%%     University, Boston, MA}}

%% %\classification{}
%% \keywords      {}

%% \author{Manuel Toharia}{
%%   address={Maryland Center for Fundamental Physics,\\  
%% Department of Physics, University of Maryland, \\
%% College Park, MD 20742, USA.}
%% }

\title{\begin{flushright}
       \mbox{\normalsize \rm UMD-PP-09-054}
         \end{flushright}
  \vskip 20pt\bf Flavor Bounds and Phenomenology in\\ the Scalar Sector of RS
Scenarios\footnote{Talk presented at SUSY09 Conference, Northeastern
University, Boston, MA}} 
\author{Manuel Toharia\\
{\it Maryland Center for Fundamental Physics,\\ Department of Physics, University of Maryland\\
College Park, MD 20742, USA}}
\date{\today}

\begin{abstract}
 
In the context of a warped extra-dimension with Standard Model fields
in the bulk, we obtain the general flavor structure of the couplings to
fermions of both the Higgs scalar and the radion graviscalar. In the
Flavor Anarchy paradigm, these couplings are generically misaligned
with respect to the fermion mass matrix and moreover the off-diagonal
couplings can be estimated parametrically as a function of fermion
masses and the observed mixing angles. One can then study the flavor
constraints and predictions arising from these couplings. 

\end{abstract}

\maketitle

%%%%%%%%%%%%%%%%%%%%%%%%%%%%%%%%%%%%%%%%%%%%
%% MAINMATTER
%%%%%%%%%%%%%%%%%%%%%%%%%%%%%%%%%%%%%%%%%%%%

Scenarios with a warped extra-dimension were introduced to address the
hierarchy problem~\cite{RS1} but one can simultaneously attack the
flavor hierarchy puzzle of the Standard Model by placing all the
fields (except the Higgs) in the Bulk. 
The observed hierarchies in masses and mixings in the fermion sector are explained
by small overlap integrals between fermion wave functions and the Higgs wave
function along the extra dimension. The electroweak
precision tests push the possible scale of new 
physics at around a few TeV \cite{EWPTmodel} and $\Delta F=2$
processes push the generic bound to be above $\sim 10$ TeV
\cite{Weiler} (see also for example
\cite{Blanke:2008zb,Casagrande:2008hr}), making it very hard to
produce and observe heavy resonances of this mass at the LHC (see also
\cite{Agashe2site} for possible ways of softening this bound and
\cite{Gedalia:2009ws} for possible limitations). 

Generically, these models contain in their spectrum two light scalars,
namely the Higgs and the radion. The Higgs scalar can arise as a
single 4D scalar localized on the TeV boundary, or as the lightest
Kaluza-Klein mode of a 5D bulk Higgs. 
The radion graviscalar can be thought of as a scalar component of the
5D gravitational perturbations, and basically it tracks fluctuations
of size of the extra-dimension (i.e. its ``radius''). 
It has recently been pointed out that in this context, one generically
expects some amount of flavor changing neutral currents (FCNC's)
mediated by both the Higgs \cite{Agashe:2009di,Azatov:2009na} and the radion \cite{Azatov:2008vm}.
The 5D spacetime we consider takes the usual Randall-Sundrum form~\cite{RS1}:
\bea 
\vspace{-.3cm}
ds^2 = \frac{1}{(kz)^2}\! \Big(\eta_{\mu\nu}
dx^\mu dx^\nu -dz^2\Big), \label{RS}
\eea
where $k$ is the curvature scale of the AdS space. 
Let's focus on the down-quark sector of a simple setup in which
we consider the 5D fermions $Q$, $D$ (containing the 4D SM
$SU(2)_L$ doublet and singlet), with 5D action
\bea
\label{fermionaction} 
\hspace{0.cm}&&
S_d\!=\!\int d^4x dz \sqrt{g} \Big[ {i \over 2}\
%\left(
\bar{Q} {\cal  D}\hspace{-.27cm}/\  Q 
%-{\cal D}_A \bar{ Q} \Gamma^A Q\right)  %\non&&\hspace{-.5cm}
+ {c_{q} \over R} \bar{ Q} {Q} + (Q\!\!\to\!\! D)
+\left(Y_d\ \bar{Q} { H} D + h.c.\right) \Big],
\eea
where $c_{q}$ and $c_{d}$ are the 5D fermion mass coefficients.
% which will determine the profile of the wavefunctions of the fermion zero modes. 
Since the Higgs $H$ is localized towards the TeV boundary, it is
useful to work with the value of the fermion zero-modes 
wavefunctions at that boundary, i.e $ f_c \equiv
\sqrt{\frac{1-2c}{1-\varepsilon^{1-2c}}}$, where $\varepsilon\approx
10^{-15}$ is the warp factor. When $c>1/2$ the dependence on the fermion
bulk mass parameter $c$ is exponentially sensitive, leading to the
explanation of mass hierarchies in this scenario.

\begin{figure}[t]
  \includegraphics[width=7.8cm,height=.3\textheight]{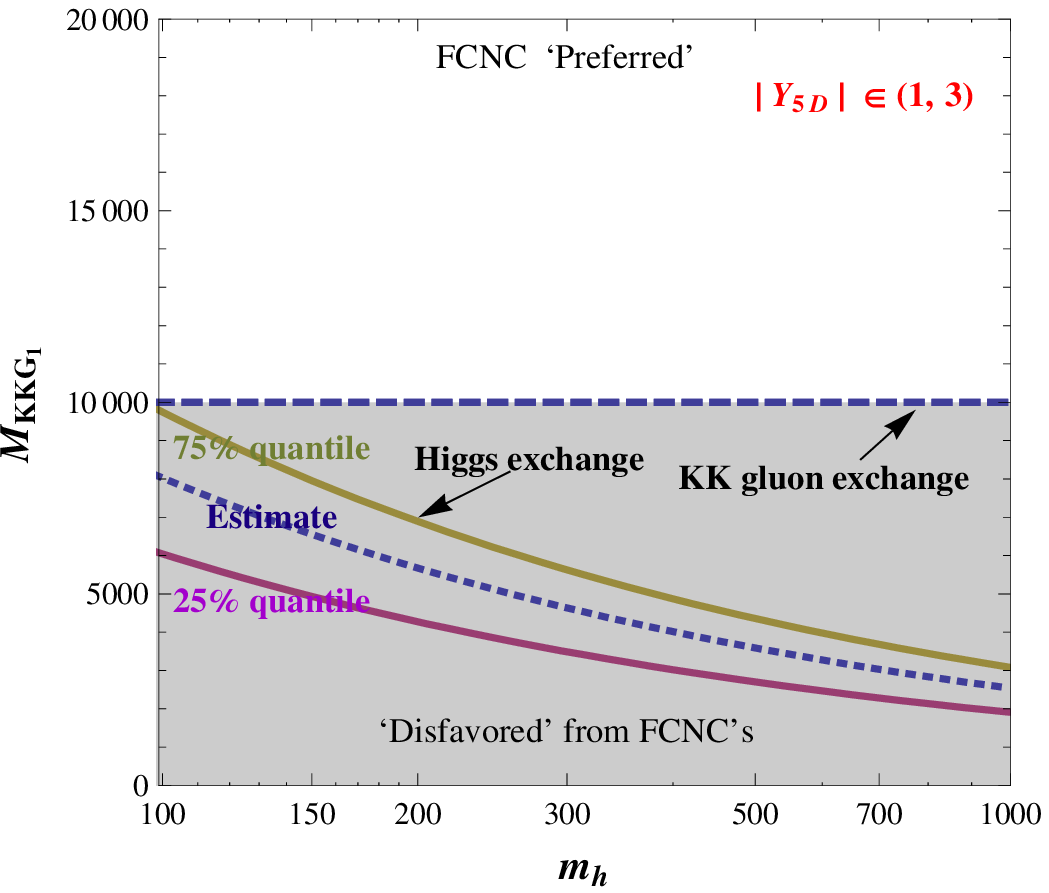}\ \  
\includegraphics[width=7.8cm,height=.3\textheight]{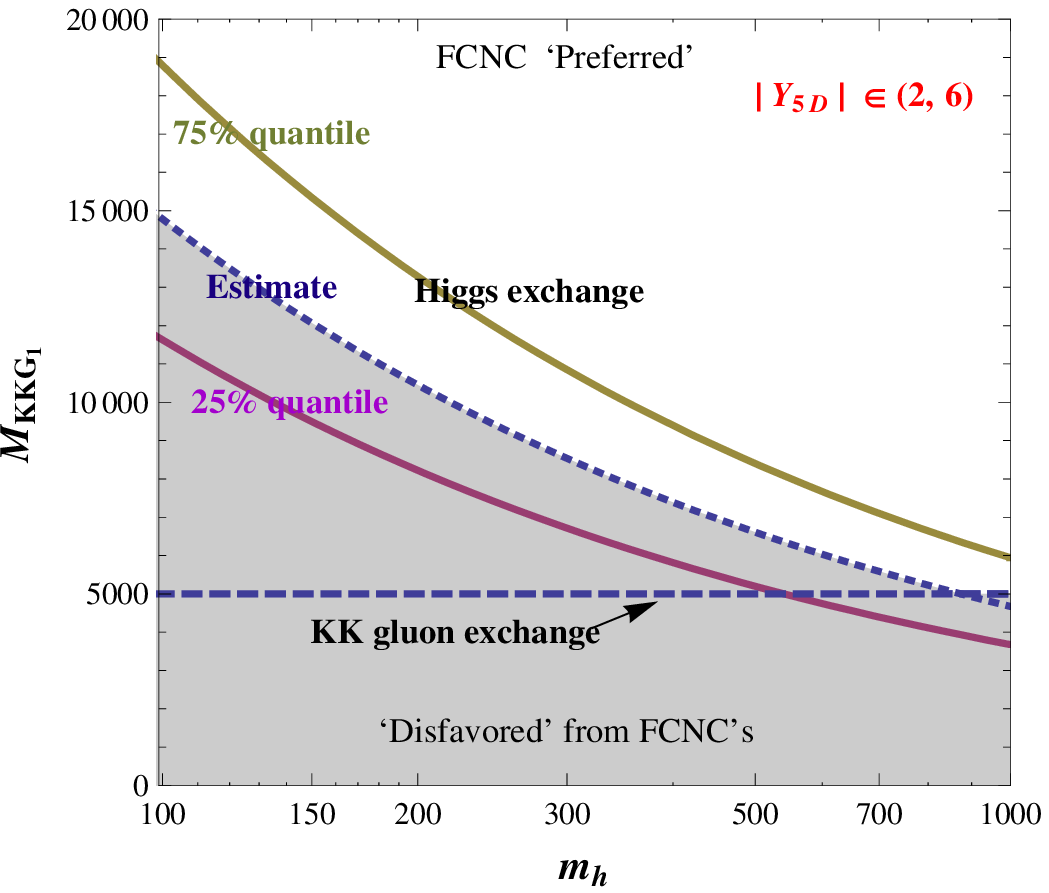}
\vspace{-.4cm}  
\caption{Generic bounds from $\epsilon_K$ in the plane ($m_h-M_{KKG1}$)
    due to Higgs exchange and KK gluon exchange, for two ranges
    of the 5D Yukawas $Y$ (left and right).}
\vspace{.8cm}
\label{higgsepsk}
\end{figure}

It turns out that generically one should expect some
misalignment between the SM fermion mass matrix and the Higgs Yukawa
coupling matrix \cite{Agashe:2009di}, irrespective of the Higgs being exactly brane 
localized or leaking into the bulk \cite{Azatov:2009na}:
%% For just one family, the shift $\Delta^d=m_d-Y_d v_4$ between mass and Yukawa
%% terms can be estimated to be
%% \bea
%% \Delta^d & \approx & m_d \frac{Y_*^2 v_4^2}{M_{KK}^2}\left[2 + {f_{c_q}^2} + f_{c_d}^2\right]
%% \eea
%% where the terms proportional to the $f$'s are subdominant (for light
%% quarks).
%the main effect coming from the diagrams: 

\begin{figure}[h!]
  \includegraphics[height=.09\textheight]{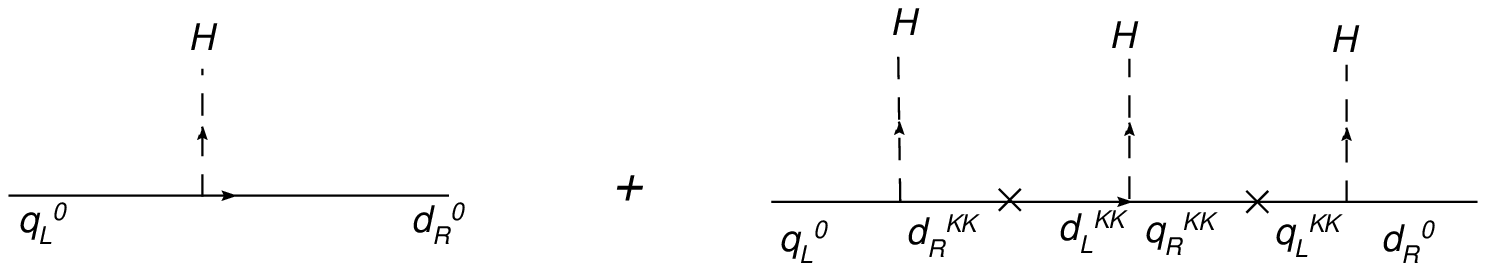}
 \label{insertion}
\end{figure}
From the second diagram, we see that when the Higgs acquires its vev,
only one term will contribute to the fermion mass, whereas three terms (from three different combinations)
will contribute to the Yukawa coupling between the physical Higgs and
two fermions, therefore potentially misaligning both matrices in
flavor space. If we parametrize the Yukawa couplings of the Higgs as 
%\begin{equation}
$\ {\cal{L}}_{HFV} = a^d_{ij}\sqrt{\frac{m^d_i m^d_j}{v_4^2}} H
\bar{d}_L^i d_R^j + h.c.$ %+ (d \leftrightarrow u)\ 
%\end{equation}
we can then estimate the size of these as
\begin{eqnarray}
a^d_{ij} \sim \delta_{ij}  -\frac{2}{3} \frac{Y^2 v_4^2}{M_{KK}^2}
\left(\begin{array}{ccc} 1 &
\lambda\sqrt{\frac{m_s}{m_d}} & \lambda^3 \sqrt{\frac{m_b}{m_d}}\\
\frac{1}{\lambda}\sqrt{\frac{m_d}{m_s}} & 1 & \lambda^2
\sqrt{\frac{m_b}{m_s}} \\ \frac{1}{\lambda^3}\sqrt{\frac{m_d}{m_b}}
& \frac{1}{\lambda^2}\sqrt{\frac{m_s}{m_b}} & 1
\end{array} \right)\label{adest}
\end{eqnarray}
where $\lambda\sim 0.22$ is the Cabibbo angle. The couplings for the
up quark sector can be found by simply replacing the down quark masses by
the up quark masses.
With these estimates, we then perform a random scan on the flavor
parameters of the setup leading to correct masses and mixings, and
obtain a distribution of the normalized Yukawa couplings
$a^{d,u}_{ij}$. In Figure \ref{higgsepsk} we show the bounds coming from $\epsilon_K$
in $\bar{K}-K$ mixing, in the plane $m_h$-$M_{KKG}$, where $M_{KKG}$ is
the first $KK$ gluon mass. On the left panel we consider smaller 5D
Yukawa values, and include a $10$ TeV bound due to KK gluon
exchange. On the right panel, we consider 5D Yukawas twice as large,
and show how this tightens the Higgs exchange bounds and reduces KK
gluon exchange tensions (although these large Yukawas put the setup on
the limits of 5D perturbativity).

\begin{figure}[t]
  \includegraphics[width=10cm,height=.32\textheight]{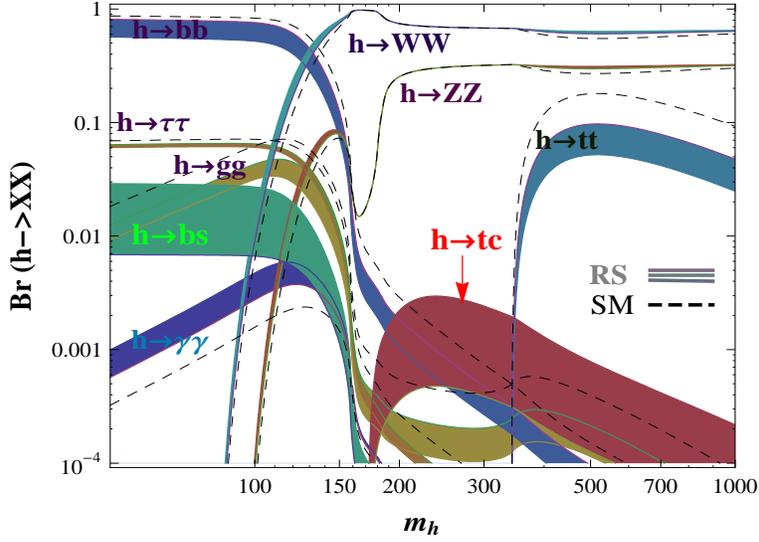}
  \caption{Branching ratios of the Higgs in the SM and in RS, with
    possible new flavor violating decays such as $h\to tc$ or $h\to bs$, for a KK
    scale of $M_{KKG1}=3.7$ TeV and $\ Y_i$ $\sim$ (1-4).}
\label{hbr}
\end{figure}

In Figure \ref{hbr} we show how the Branchings of the Higgs change
because of the new flavor violating couplings. There is a generic suppression in the
diagonal couplings (specially to tops and bottoms) which results in
lower Branchings to fermions. Interestingly this results in an
enhanced Branching into photons, although the overall Higgs production at the
LHC will be also be generically suppressed in the gluon fusion channel. Nevertheless
one expects that some of these features will be observed at the LHC. Then one
could probe for exotic decays such as $h\to tc$, if kinematically
allowed (this will be true for the radion too).

\begin{figure}[t]
  \includegraphics[width=10cm,height=.32\textheight]{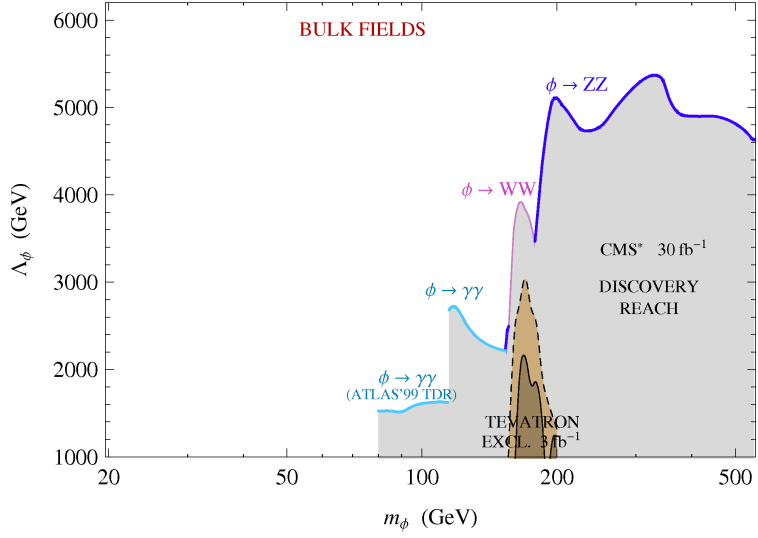}
  \caption{LHC discovery reach for the radion using ``translated'' Higgs projections from CMS
(and ATLAS in the lower mass region) for $30 fb^{-1}$ of luminosity.}
\label{radionreach}
\end{figure}

The radion graviscalar can be parametrized as a scalar perturbation
of the  metric:
\bea
\hspace{-1cm} ds^2  \label{metricpert}
&=& \left(\frac{R}{z} \right)^2(e^{-2F}\eta_{\mu\nu} dx^\mu dx^\nu - (1+2F)^2 dz^2)
\eea
Demanding that the perturbed metric solves the Einstein equation and
that the Radion field is canonically normalized, we get $\ F
\ =\ \frac{r(x)}{\Lambda_r} \frac{z^2}{R'^2}\ $
where $r(x)$ is the corresponding canonically normalized radion graviscalar with its
associated interaction scale
$\Lambda_r=\sqrt{6}\ \varepsilon\ M_{Pl}$, where again $\varepsilon
\sim 10^{-15}$. This scale is of TeV size and therefore the radion can
have interesting collider phenomenology \cite{GRW,Korean}. In Figure
\ref{radionreach}, the LHC discovery reach for the radion is plotted
as a function of its mass and the interaction scale $\Lambda_r$.

\begin{figure}[t]
  \includegraphics[width=10cm,height=.32\textheight]{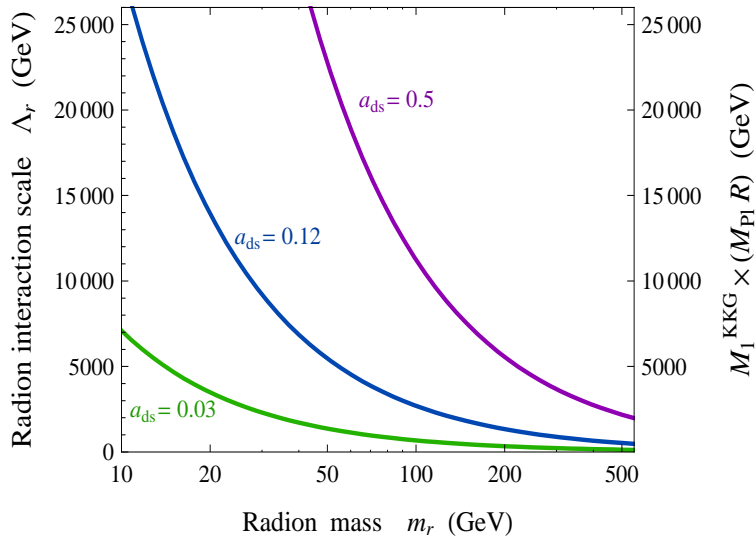}
 \caption{Generic bounds from $\epsilon_K$ in the plane ($m_r-M_{KKG1}$)
    due to tree-level radion exchange.} 
\label{radepsk}
\end{figure}

Similarly to the
Higgs, its couplings to fermions can be parametrized as
%\bea
$\ {\cal{L}}_{rFV} = {1\over\Lambda_r} \tilde{a}^d_{ij}\sqrt{m^d_i
  m^d_j}\ \  { r(x)} { \bar{d}_L^{\,i} d_R^j}\  + h.c.\ $
%\eea
and these can have off-diagonal entries, which we estimate as
\cite{Azatov:2008vm}:  
\begin{eqnarray} 
\vspace{-.3cm}
\tilde{a}^d_{ij} \sim\left(\begin{array}{ccc} 
(c_{q_1}-c_{d_1}) & (c_{q_1}-c_{q_2})\lambda \sqrt{\frac{m_s}{m_d}}
  &G(c_{q_i}) \lambda^3 \sqrt{\frac{m_b}{m_d}} \\ 
 (c_{d_1}-c_{d_2}) \frac{1}{\lambda}\sqrt{\frac{m_d}{m_s}} & \left(c_{q_2}-c_{d_2}\right) &  (c_{q_2}-\frac{1}{2})\lambda^2
  \sqrt{\frac{m_b}{m_s}} \\
F(c_{d_i}) \frac{1}{\lambda^3}\sqrt{\frac{m_d}{m_b}} &  (c_{d_2}-c_{d_3})\frac{1}{\lambda^2}\sqrt{\frac{m_s}{m_b}} & (\frac{1}{2}-c_{d_3}) 
\end{array} \right)
\end{eqnarray}
where $F$ and $G$ are some combination of $c_i$'s and extending to the up quark
sector is immediate. 
Generic bounds can again be obtained, in particular those coming from
$\epsilon_K$ in $\bar{K}-K$ mixing, as shown in Figure \ref{radepsk}, where it is seen that a
very light radion is highly disfavored. After discovering the radion
at the LHC (say from $r\to ZZ$ decays), it would be important to
search for exotic decays such as $r\to tc$. 

We conclude with the message that the two potentially light scalars of
these scenarios, which presumably could be discovered at the LHC have
some amount of Flavor Violating couplings, and probing these will be a
very important probe on the origin of flavor and of these setups in general.
%\vspace{-.5cm}

%%%%%%%%%%%%%%%%%%%%%%%%%%%%%%%%%%%%%%%%%%%%%%%%
%% BACKMATTER
%%%%%%%%%%%%%%%%%%%%%%%%%%%%%%%%%%%%%%%%%%%%%%%%

{\bf Acknowledgments}
%\begin{theacknowledgments}

I would like to thank my collaborators Aleksandr Azatov and Lijun Zhu
as well as Kaustubh Agashe for discussions and comments. 
%\end{theacknowledgments}

%\vspace{-.5cm}

\end{document}